# The DREAMS Project: Improving the Intensive Care Patient Experience with Virtual Reality

Triton L Ong, MA[1], Matthew M Ruppert, BS[2,5], Parisa Rashidi, PhD[3,5], Tezcan Ozrazgat-Baslanti, PhD[,2,5], Azra Bihorac, MD MS[2,5], Marko Suvajdzic, PhD[4]

[1]Department of Psychology, University of Florida, Gainesville, FL, USA

[2]Department of Medicine, University of Florida, Gainesville, FL, USA

[3]Department of Biomedical Engineering, University of Florida, Gainesville, FL, USA

[4]Digital Worlds Institute, University of Florida, Gainesville, FL, USA

[5]Precision and Intelligent Systems in Medicine (Prisma[P]), University of Florida, Gainesville, FL, USA

**Correspondence can be sent to:** Azra Bihorac, MD, MS, FASN, FCCM, Department of Medicine, Division of Nephrology, Hypertension, & Renal Transplantation, 1600 SW Archer Road, PO Box 100224, Communicore Building, Room CG-98, Gainesville, FL 32610-0224. Office Phone (352) 273-9009; Cell Phone: (352) 870-0090; Fax: (352) 392-5465; E-mail: abihorac@ufl.edu. Reprints will not be available from the author(s).

**Key Words**: virtual reality, intensive care unit, patient experience, delirium, anxiety, depression.

**Conflicts of Interest and Source of Funding:** This work is supported by NSF CAREER 1750192 (PR), NIH/NIBIB 1R21EB027344-01 (PR), NIH/NIGMS RO1 GM-110240 (MR, PR, TOB, AB), NIH/NIGMS P50 GM111152 (AB, TOB), and Davis Foundation – University of Florida (MR).  The authors declare that they have no competing interests.

**Ethics Approval and Consent to Participate:** All procedures were reviewed and approved by UF IRB-02. All participants provided informed consent prior to participation.

**Availability of Data and Material:** The datasets generated and/or analyzed during the current study are not publicly available to protect patient confidentiality but are available from the corresponding author on reasonable request.

**Trial registration:** The trial was registered on December 29, 2017 with ClinicalTrials.gov with the identifier: NCT03385993.

**Authors' Contributions**: The authors Triton Ong (TO) and Matthew Ruppert (MR) contributed equally to this manuscript. TO contributed to study development, analysis, and writing. MR contributed to study development, managed data collection, performed statistical analysis, and contributed to writing. Parisa Rashidi coordinated recruitment and data collection. Tezcan Ozrazgat-Baslanti assisted in data management and analysis. Marko Suvajdzic oversaw VR technologies and session procedures. Azra Bihorac supervised medical conduct and ICU protocol. All authors read and approved the final manuscript.

**Acknowledgements:** The authors would like to thank Sherry Brown and George Omalay for their valuable assistance in planning and conducting this research. The authors would also like to thank research coordinators (Haleh Hashemi, Julie Cupka, and Laura Velez) and research assistants (Ria Bhaskar, Ryan Cherico, Elizabeth Ingersent, and Emilie Pearson) for their assistance during data collection and analysis. The authors would like to thank *RelaxVR* for allowing the use of the screenshot depicted in Fig1b.

**Abstract**

- **Purpose:** Preliminarily evaluate the feasibility and efficacy of using meditative virtual reality (VR) to improve the hospital experience of intensive care unit (ICU) patients.

- **Methods:** Effects of VR were examined in a non-randomized, single-center cohort. Fifty-nine patients admitted to the surgical or trauma ICU of the University of Florida Health Shands Hospital participated. A Google Daydream headset was used to expose ICU patients to commercially available VR applications focused on calmness and relaxation (*Google Spotlight Stories* and *RelaxVR*). Sessions were conducted once daily for up to seven days. Outcome measures included pain level, anxiety, depression, medication administration, sleep quality, heart rate, respiratory rate, blood pressure, delirium status, and patient ratings of the VR system. Comparisons were made using paired t-tests and mixed models where appropriate.

- **Results**: The VR meditative intervention was found to improve patients' ICU experience with reduced levels of anxiety and depression; however, there was no evidence suggesting that VR had any significant effects on physiological measures, pain, or sleep.

- **Conclusion:** The use of VR technology in the ICU was shown to be easily implemented and well-received by patients.

## Introduction

Patients' intensive care unit (ICU) stays are often traumatic. Prolonged immobility and sedation during treatment can lead to ICU-acquired weakness (ICUAW) [1]. Disrupted sleep, long stays, and extended periods of pain put ICU patients at greater risk for delirium-related mortality [2]. After ICU discharge, 50%-70% of patients exhibit persistent dysfunction, weakness, and post-traumatic symptoms that can have indefinite impacts on the patient's finances, independence, and daily life [3].

Many of these ICU-related complications are not the direct result of illness, injury, or treatment. Critical care professionals have raised attention to modifiable aspects of the ICU to improve patient recovery experience [4]. Early regular exercise programs have shown promising results for preventing ICUAW [5]. Clinical guidelines for delirium prevention emphasize strategies to orient patients, manage pain, control noise and light, and promote good sleep [6-8]. Although modifiable risks have been identified, there are few feasible strategies for mitigating these risks within the ICUs given the limited time and resources.

We hypothesize that virtual reality (VR) can provide a platform for controlled, scalable, and effective environmental manipulation in the ICU. VR uses a head mounted display to deliver immersive video and audio that enables interaction through tracking head, hand, and/or body movement [9]. VR has been praised for mitigating some of the limitations of traditional therapies. VR experiences can help users feel safer, more in control, and more comfortable than in-person outpatient therapy through direct visualization without the stress of real stimuli [10]. Exposure therapy in VR was demonstrated to be as effective as standard in-situ treatment while being perceived as more tolerable by patients [11]. Preoperative VR relaxation has been shown to reduce anxiety and stress in child and adult patients [12, 13]. Severe burn victims reported less pain when VR relaxation was used during wound debridement [14]. Medical therapy with VR has been generally efficacious and accepted across a variety of treatment contexts. It remains important to expand VR applications towards improving the patient experience [15].

Recommendations for optimal ICU settings encourage early exercise, comfortable ambiance, pain management, and good sleep. In addition to the environmental adjustments provided by the healthcare team, VR may provide a system in which some of these recommendations can be enhanced. The purpose of this interdisciplinary study was to evaluate feasibility of VR relaxation therapies for ICU patients.

## Materials and Methods

**Participants and Setting**

This study was conducted on a single-center cohort of patients admitted to the surgical or trauma ICU at University of Florida (UF) Health Shands Hospital, a large academic quaternary care facility in the Southeastern United States. Participants were ≥18 years, negative for delirium, not in contact isolation (e.g., not infectious or at high risk for infection), likely to remain in the ICU for ≥48 hours, non-intubated, and lacking conditions which might limit head or neck movement. All study procedures were performed in the patients' ICU rooms. The study was approved by the UF Institutional Review Board (#IRB201703107).

**Materials**

The VR system consisted of a smartphone placed in a Google Daydream (vr.google.com/daydream) headset and a pair of Bluetooth headphones (Fig1a). The equipment was lightweight (< 1 pound) and easy to adjust. We used *Google Spotlight Stories' Pearl* (atap.google.com/spotlight-stories) as an initial orientation to VR and *RelaxVR* (www.relaxvr.co; Fig1b) to provide patients with a calm immersive scene (e.g., rolling waves on a beach) with voice guided meditation that promoted breath control and relaxation. The headset and earphones were affixed with disposable sanitary covers. The headset, smartphone, controller, VR applications, and earphones were collectively referred to as the Digital Rehabilitation Environment Augmenting Medical System (DREAMS).

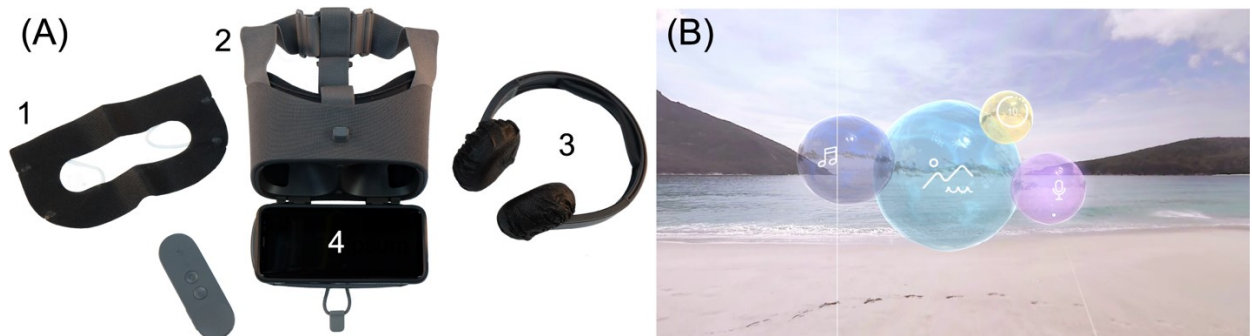

**Fig 1.** (a) DREAMS Equipment: 1, VR sanitary mask; 2, Google Daydream headset; 3, Bluetooth headphones with sanitary covers; 4, Android smartphone. (b) RelaxVR menu screenshot.

**Dependent Measures**

The primary dependent measures of this study were participants' pain, sleep quality, affect, delirium, and responses to using DREAMS. Pain was measured with the Defense and Veterans Pain Rating Scale (DVPRS) [16], sleep quality with the Richards-Campbell Sleep Questionnaire (RCSQ) [17], affect with the Hospital Anxiety and Depression Scale (HADS) [18], delirium status with the Confusion Assessment Method for the Intensive Care Unit (CAM-ICU) [19], and patients' qualitative responses to DREAMS with structured interviews (Supplement A). Each patient's heart rate (HR), respiration rate (RR), blood pressure (BP), and medication records were utilized to evaluate if the VR sessions had any effect on physiology and pain.

The CAM-ICU, DVPRS, HR, RR, BP, and medication records were recorded by healthcare staff during normal care. Records were retrieved from the UF Integrated Data Repository after sessions were concluded. The RCSQ, HADS, and DREAMS questionnaires were administered by study staff during sessions.

**Session Procedures**

Study staff administered the RCSQ and HADS on the first day of the study to establish baseline measures. Participants were then fitted with DREAMS and exposed to *Pearl* (5 min) to demonstrate the format of VR. Study staff then initiated a 5-20 minute guided meditation for breath control and progressive relaxation using the *RelaxVR* app. Once the session was completed, study staff removed the headset and interviewed participants with open-ended questions about their experience. At the end of the session, researchers asked participants to revisit the relaxation techniques provided by *RelaxVR* whenever they felt it could help.

Participants received up to seven sessions, each at least 24 hours apart. *Pearl* was only shown during the initial session, and subsequent sessions occurred in an otherwise identical manner.

**Data Analysis**

Results were summarized as frequencies and percentages for categorical variables, mean and standard deviation for normally distributed variables, and median and interquartile ranges for non-normal continuous variables. Paired t-tests with adjustments made for multiple comparisons were used to compare pre- and post-session numerical values. Mixed models were constructed to examine the changes in DVPRS, HR, RR, BP, opioid medication dosage, 'pro re nata' (PRN) opioid medication dosage, PRN opioid medication frequency, RCSQ, and HADS across study days taking into account the correlation within the same subject's measurements. As a sensitivity analysis for measures collected multiple times per day, we constructed models comparing pre- and post-DREAMS session values within one, two, four, six, eight, and twelve hours of the DREAMS session. Dosages of opioids were converted to oral morphine milligram equivalents prior to analysis. Medications received during an operation were excluded from the analysis. Statistical features were extracted from time series physiological data including min, max, variance, and mean across study days for all time intervals. All significance tests were two-sided with α < 0.05 considered statistically significant. Statistical analyses were performed with R v.3.6.

## Results

**Participants**

A total of 59 participants were recruited (Fig2). Thirteen participants did not complete the study due to emergent surgery or discharge from the ICU. The remaining 46 participants received either one (N = 17), two (N = 17) or three to seven (N = 12) DREAMS sessions. Participants were generally older ($M$ = 50 years, $SD$ = 18) and male (65%) (Table 1). The median hospital stay for participants was 11 days.

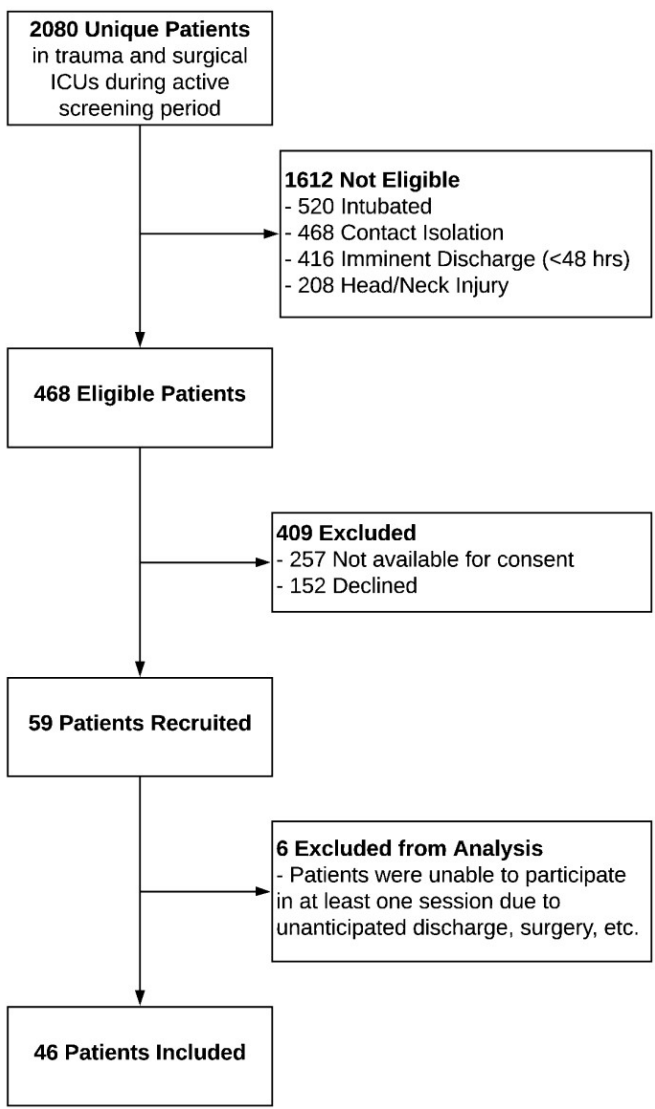

**Fig 2.** Screening Consort Diagram.

**Table 1**. Patient Demographics

| Variables | N=46 |
|---|---|
| **Baseline Characteristics** | |
| Age in years, mean (SD) | 50 (18) |
| Sex (male), n (%) | 30 (65%) |
| Ethnicity, n (%) | |
|   Black | 8 (18%) |
|   Other ethnicity | 2 (4%) |
|   White | 36 (78%) |
| Admission Type, n (%) | |
|   Emergent | 28 (61%) |
|   Routine elective | 18 (39%) |
| **Hospital Outcomes** | |
| Length of stay (days), median (25th-75th) | 11 (7,23) |
| Discharge Disposition, n (%) | |
|   Death | 1 (2%) |
|   To another hospital | 3 (7%) |
|   To home | 16 (35%) |
|   To homecare | 13 (28%) |
|   To other lower facility | 13 (28%) |

**Pain**

The DVPRS is a self-reported visual analog scale that ranges from 0 ("no pain") to 10 (so painful that "nothing else matters"). No statistically significant relationship was found between study day and DVPRS at any time point (p > 0.05; FigS1).

Despite no statistically significant improvement in DVPRS, 81% of patients agreed or strongly agreed with the statement, "I feel that I experienced less pain yesterday because of the DREAMS" (FigS2). This discrepancy between patients' objective (DVPRS) and subjective (DREAMS questionnaire) pain ratings could be due to confirmation bias, demand characteristics, or subtle effects of VR not identified in this study and requires further investigation.

The dosage and frequency of opioid medications decreased over time at a rate of 12.9 (95% CI 21.7, 4.03) oral morphine milligram equivalents (MMEs) per study day. No statistically significant changes were found when comparing dosage or frequency before and after any intervention. Nonetheless, the observed decreases in PRN opioid dosages may be clinically significant with an average decrease from 54.8 MME after the first intervention to 11.5 MME after the third intervention (TableS1).

**Sleep**

The RCSQ is a series of questions about last night's rest that patients scored from 0 to 100 (higher indicates better sleep). Participants' RCSQ score improved by 4.56 (95% CI 1.06, 8.06) points each study day (FigS3), however there was no statistically significant difference observed when comparing successive nights sleep or baseline sleep quality to a given study day.

**Affect**

The HADS is a scale for patients to estimate their current anxiety and depression. A sub-score of 0-7 is considered normal affect, 8-10 borderline, and 11-21 abnormal. We compared participants' HADS' sub-scores before their first DREAMS exposure to just before their second and third exposures. There was no statistically significant change in anxiety or depression before the second session, however there was a statistically significant change in anxiety (estimate = -2.17, 95% CI: -4.23,-0.106) and depression (estimate = -1.25, 95% CI: -2.37,-0.129) from before the first exposure to before the third exposure (Fig3). Ten of thirteen patients with borderline depression improved to normal during the study. No patients transitioned to a worse depression classification during the study. Four of ten patients with abnormal anxiety improved, with two of those patients reaching normal range during the study. Five of the eleven patients with borderline anxiety improved to normal during the study. Three patients experienced an increase in anxiety during the study.

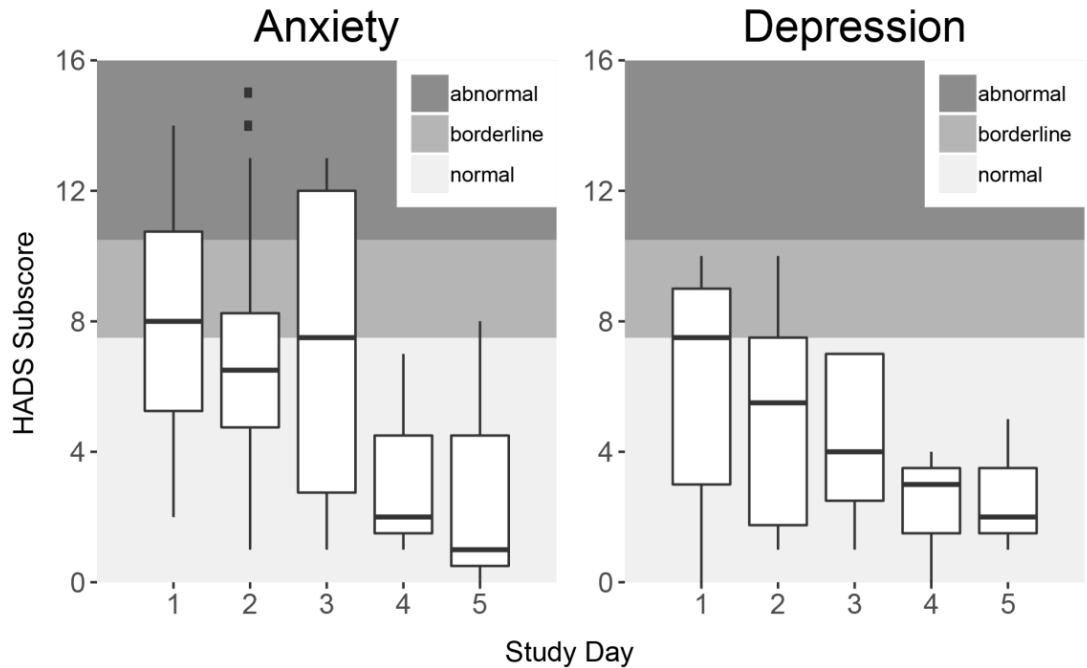

**Fig 3.** Hospital Anxiety and Depression (HADS) subscores for DREAMS participants

**Delirium**

Of the 46 subjects that participated in a DREAMS session, 13 were delirious for at least one day during their admission. Seven participants were delirious prior to the study but recovered before enrollment and remained non-delirious until discharge. The other six patients became delirious after completing the DREAMS study and were diagnosed an average of 84 hours after their final DREAMS session.

**Vital Signs**

Patients' systolic BP, diastolic BP, mean arterial pressure, HR, or RR were compared at one, two, four, six, eight, and twelve hours before and after each DREAMS session. No statistically significant differences were observed in pre- vs post-session, minimum, maximum, mean, or variability at any time interval or session number.

**Participants' Reactions to DREAMS**

Participants were asked to rate and discuss how much they agreed with statements about their use of DREAMS. They agreed that DREAMS was comfortable (*Comfort*), enjoyable (*Enjoyment*), helped them better manage their pain (*Pain*), and that they thought about DREAMS outside of sessions (*Reflection*). However, participants were mixed on whether DREAMS helped them sleep better (*Sleep*) (Fig4). Transcripts of audio recordings were analyzed qualitatively to identify themes in participants' responses.

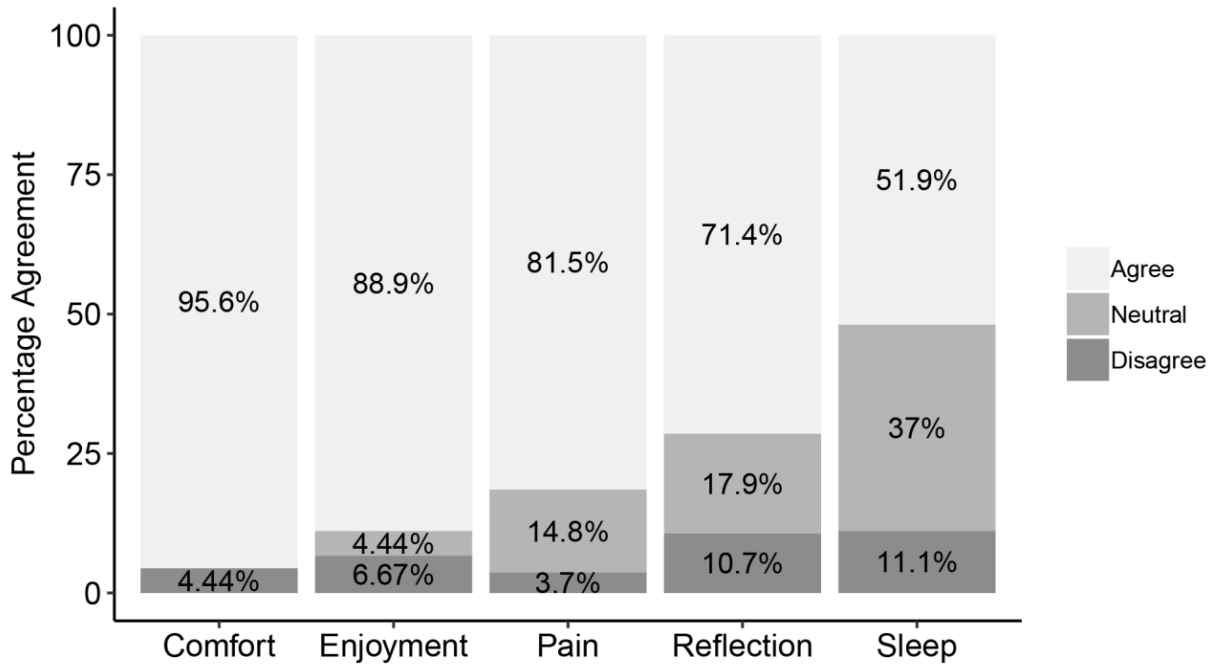

**Figure 4.** Participants' reaction to the DREAMS: Comfort = "I thought using the DREAMS was comfortable," Enjoyment = "I liked the experience of using the DREAMS," Pain = "I feel that I experienced less pain yesterday because of the DREAMS," Reflection = "I found myself thinking about the DREAMS after the session was over," Sleep = "I feel that I slept better last night because of the DREAMS"

**Novelty of VR.** Although only one participant had prior experience with VR, the enjoyment of DREAMS was universal. Participants were often observed smiling, laughing, and giving positive remarks during use of DREAMS. Several participants recommended wider deployment of DREAMS. "[Other ICU patients] would be crazy if they didn't [want to try DREAMS]…she was telling you how to breathe, and I could see how that could be beneficial for us in here, or for anybody." Participants enjoyed DREAMS enough to inquire about purchasing. "I was going to ask you if I could buy one," "Is this on iPhone? You should put it on an app!" Participants also volunteered feedback for improvements of DREAMS. Most comments involved better visuals and audio for improved immersion. "The waves were good, don't get me wrong…But to get the authenticity of it all, you need [seagulls] just in the distance." Participants were generally interested in the novelty of VR and expressed excitement about its use in the ICU.

**Emotionally evocative.** The immersive virtual environments depicted in *Pearl* and *RelaxVR* tended to evoke nostalgic feelings. One participant noted, "[the rocks] remind me of the shores of Maine" while another stated that the beach scene made her think of family vacations she would like to plan after discharge. Some emotional reactions to *Pearl* were unexpected. *Pearl* is a short VR film about a family told from the inside of their family car. Several participants reported feeling fearful of the depiction. "Not safe driving…Sure tells you what not to do, when [the father] jumped into the back seat. He's stupid." Another participant noted sarcastically, "It's funny, he's driving while playing the guitar. Seems totally safe." One participant experienced an especially negative emotional reaction to *Pearl* due to a previous experience of a loved one lost in a car accident. The immersive properties of VR can be a powerful tool for health promotion. It is important, however, that researchers, clinicians, and developers be aware of potential negative reactions that can be particularly strong in VR.

**Relaxing.** Participants would often vocalize statements about relaxation and drift into sleep during VR sessions. For some participants, DREAMS provided a welcomed feeling of privacy and escape from their ICU room. Participants noted that "I liked that I could enjoy it [by myself]" and "[VR] is better than the TV because you're there [the beach]." The DREAMS provided participants with an isolated visual and auditory environment to focus on their relaxation and breathing, which may have provided an escape or distraction from their uncomfortable but necessary recovery situation.

**Technical frustrations.** Sometimes participants were provided with suboptimal VR experiences due to errors with the equipment or software. Headsets that shifted during sessions would allow light from the ICU to bleed into the screen and disrupt the viewing experience. A blinking notification light became activated during one participant's session, who later noted feeling dizzy as a result. Technical and procedural difficulties can contribute to participant frustration and should be minimized with training and preparation to provide the best experience possible. It will also be important to tailor VR equipment for the unique demands of the ICU. Most commercially available VR apps require the user to walk or rotate to fully engage with the VR experience—a potentially uncomfortable or unsafe prospect for many ICU patients.

**Temporary effects.** Participants seemed aware that the physiological effects of DREAMS were negligible. The DREAMS was reported to have distracted from pain, but only during VR exposure. Several participants noted that they received medication that affected their sleep and made it difficult to say if DREAMS contributed. "Unfortunately, it was probably the melatonin…I'm sure [DREAMS helped me sleep] because I did not have the same problem." Participants were unsure if DREAMS helped with pain, mostly due to the critical nature of their status. "I can't tell [if DREAMS helped with my pain] because yesterday was pretty painful." Another participant noted, "I mean, my leg pain was real bad, so I can't really put it on [DREAMS]." Despite this, participants still reported enjoying DREAMS. When asked if DREAMS helped with his pain, one participant responded, "Not really, but it's a good part of my day."

**Discussion**

We demonstrated initial feasibility for use of VR relaxation in ICU patients. Despite finding no clinically or statistically significant effects on physiology, pain, or sleep, participants overwhelmingly enjoyed the VR experiences provided by DREAMS. Our results show that ICU patients are eager to undergo the stimulation that VR provides and may serve as a welcomed distraction from unavoidable discomforts of their medical situations. Collectively, these results show VR to be a promising option to help improve the ICU patient experience.

The ICU is a busy, noisy setting in which patients of the greatest need are closely monitored by highly trained staff. Proposed additions to the ICU must be effective, simple, and affordable. DREAMS equipment was easy to set up, intuitive to operate, and enjoyed by participants. Each component of DREAMS is widely available, and most people are already familiar with smartphone interfaces. Furthermore, DREAMS can be quickly and easily sanitized with common medical disinfectant wipes. Combined with participants' enjoyment of the VR, this makes DREAMS and similar systems remarkably portable and an ideal candidate for deployment in the ICU.

The primary goal of this study was to assess feasibility of VR for ICU patients' experience. While we did not find clinically or statistically significant effects in health outcomes such as pain, vital signs, or sleep, it should be noted that participants only received 5-20 minutes of VR exposure each day and that granularity of vital signs data was limited. It seems unlikely that 5-20 minutes of VR exposure would produce large effects in a critical care environment, but this remains an important topic to evaluate in future research. Previous research has shown that 40 minutes of VR exposure repeated across three months has been implemented with positive results [15]. There remains good reason to hypothesize that VR can help patients better manage stress and discomfort in the ICU [20].

Participants in this study may represent selection bias. We approached ICU patients who were conscious, not intubated, not in isolation, and not already delirious. While VR would not be helpful for the unconscious or severely delirious, ICU patients who are otherwise awake should be included in future research as they are at the greatest risk for developing ICUAW and delirium. It will be important for researchers, clinicians, developers, and ICU survivors to collaborate in the design of VR equipment and software specific to the ICU patient experience—including those who may be immobile, intubated, or in contact isolation.

ICU patients are likely to experience unease and uncertainty in their recovery. These patients are under constant observation and receive the best medical care available. However, the vast majority of their time in the ICU is spent in prolonged discomfort and sedentary in an austere environment. VR technologies are relatively affordable, increasingly easy to use, and enjoyed by patients. Therapies in VR can be tailored to the needs of ICU patients to help improve bodily control and pain management, reduce stress, and provide a welcome distraction from the uncomfortable nature of their current condition.

## Conclusion

A virtual reality meditative intervention was found to improve patients' experiences in the ICU by reducing patients' anxiety and depression; however, there was no evidence suggesting that VR had any significant effects on vital signs, pain, or sleep. The use of VR in the ICU was shown to be easily implemented and well-received by patients. The DREAMS project demonstrates that interdisciplinary collaborations between clinical researchers, artists, engineers, and psychologists can implement emerging technologies to improve patients' experiences in the ICU.

**References**

1. Zorowitz RD, (2016) ICU-Acquired Weakness: A Rehabilitation Perspective of Diagnosis, Treatment, and Functional Management. Chest 150: 966-971
2. Kalabalik J, Brunetti L, El-Srougy R, (2014) Intensive care unit delirium: a review of the literature. J Pharm Pract 27: 195-207
3. Svenningsen H, Langhorn L, Ågård AS, Dreyer P, (2017) Post-ICU symptoms, consequences, and follow-up: an integrative review. Nurs Crit Care 22: 212-220
4. Wilson ME, Beesley S, Grow A, Rubin E, Hopkins RO, Hajizadeh N, Brown SM, (2019) Humanizing the intensive care unit. Crit Care 23: 32
5. Wischmeyer PE, San-Millan I, (2015) Winning the war against ICU-acquired weakness: new innovations in nutrition and exercise physiology. Crit Care 19 Suppl 3: S6
6. Mistraletti G, Pelosi P, Mantovani ES, Berardino M, Gregoretti C, (2012) Delirium: clinical approach and prevention. Best Pract Res Clin Anaesthesiol 26: 311-326
7. Brummel NE, Girard TD, (2013) Preventing delirium in the intensive care unit. Crit Care Clin 29: 51-65
8. Barr J, Fraser GL, Puntillo K, Ely EW, Gélinas C, Dasta JF, Davidson JE, Devlin JW, Kress JP, Joffe AM, Coursin DB, Herr DL, Tung A, Robinson BR, Fontaine DK, Ramsay MA, Riker RR, Sessler CN, Pun B, Skrobik Y, Jaeschke R, Medicine ACoCC, (2013) Clinical practice guidelines for the management of pain, agitation, and delirium in adult patients in the intensive care unit. Crit Care Med 41: 263-306
9. Gigante MA, Jones H, Earnshaw RA (1993) Virtual reality systems. Academic Press, London
10. MC S, RI R, RA F (2011) Virtual reality supporting psychological health. In: S B, Lc J (eds) Advanced Computational Intelligence Paradigms in Healthcare 6. Springer, Berlin, pp. 13-29
11. Morina N, Ijntema H, Meyerbröker K, Emmelkamp PM, (2015) Can virtual reality exposure therapy gains be generalized to real-life? A meta-analysis of studies applying behavioral assessments. Behav Res Ther 74: 18-24
12. Ryu JH, Park SJ, Park JW, Kim JW, Yoo HJ, Kim TW, Hong JS, Han SH, (2017) Randomized clinical trial of immersive virtual reality tour of the operating theatre in children before anaesthesia. Br J Surg 104: 1628-1633
13. Ganry L, Hersant B, Sidahmed-Mezi M, Dhonneur G, Meningaud JP, (2018) Using virtual reality to control preoperative anxiety in ambulatory surgery patients: A pilot study in maxillofacial and plastic surgery. J Stomatol Oral Maxillofac Surg 119: 257-261
14. Faber AW, Patterson DR, Bremer M, (2013) Repeated use of immersive virtual reality therapy to control pain during wound dressing changes in pediatric and adult burn patients. J Burn Care Res 34: 563-568
15. Dascal J, Reid M, IsHak WW, Spiegel B, Recacho J, Rosen B, Danovitch I, (2017) Virtual Reality and Medical Inpatients: A Systematic Review of Randomized, Controlled Trials. Innov Clin Neurosci 14: 14-21
16. Buckenmaier CC, Galloway KT, Polomano RC, McDuffie M, Kwon N, Gallagher RM, (2013) Preliminary validation of the Defense and Veterans Pain Rating Scale (DVPRS) in a military population. Pain Med 14: 110-123
17. Richards KC, O'Sullivan PS, Phillips RL, (2000) Measurement of sleep in critically ill patients. J Nurs Meas 8: 131-144
18. Zigmond AS, Snaith RP, (1983) The hospital anxiety and depression scale. Acta Psychiatr Scand 67: 361-370
19. Ely EW, Margolin R, Francis J, May L, Truman B, Dittus R, Speroff T, Gautam S, Bernard GR, Inouye SK, (2001) Evaluation of delirium in critically ill patients: validation of the Confusion Assessment Method for the Intensive Care Unit (CAM-ICU). Crit Care Med 29: 1370-1379
20. Villani D, Preziosa A, Riva G (2006) Coping with stress using virtual reality: A new perspective. In: Editor (ed)^(eds) Book Coping with stress using virtual reality: A new perspective. Interactive Media Institute, City, pp. 25-32

# Supplementary Material

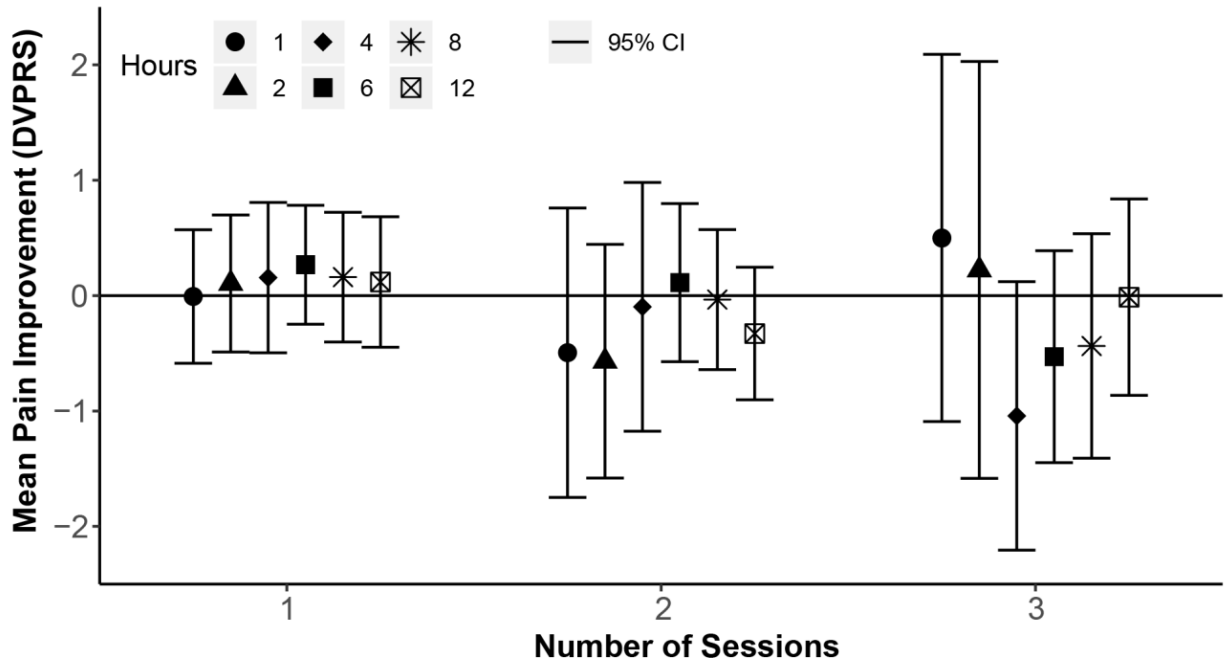

**FigS1.** Mean DVPRS pain improvement before and after VR exposure.

## "I feel that I experienced less pain yesterday because of the DREAMS"

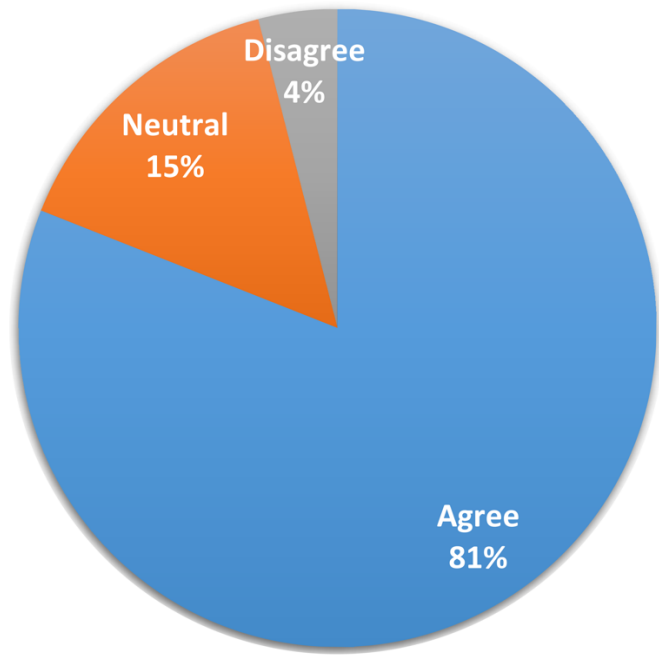

**FigS2.** Patients' perceptions of how DREAMS decreased their pain

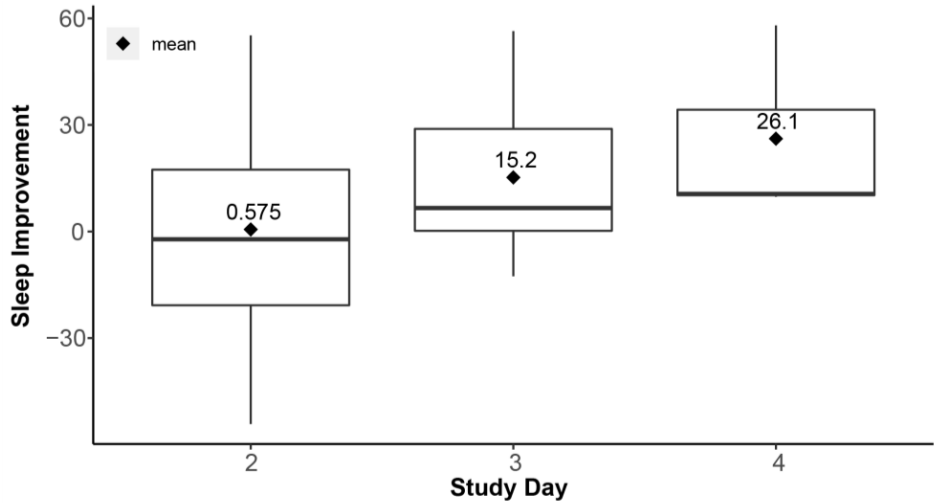

**FigS3.** Sleep Improvement over time compared to baseline

**Table S1.** PRN mg of oral morphine equivalent administered 24 hours pre-post sessions.

| Session Number | Estimated Decrease (95% Confidence Interval) |
|---|---|
| 1 | 54.8 (-41.5, 151) |
| 2 | 31.4 (-8.91, 71.7) |
| 3 | 11.5 (-4.09, 27.1) |

**Supplement A: DREAMS Interview Procedures**

After completion of the virtual reality (VR) experience, a brief interview was performed to evaluate subjects' perceptions of the DREAMS experience. The interview was audio recorded for participants that consented to audio recording.

The interview began with the following set of instructions:

"Please rate your agreement with the following statements on a scale from one to five, with one being 'Strongly Agree' and five being 'Strongly Disagree'."

1. "I liked the experience of using the DREAMS."
2. "I thought using the DREAMS was comfortable."
3. "I found myself thinking about DREAMS after the session was over."
4. "I feel that I slept better last night because of the DREAMS."
5. "I feel that I experienced less pain yesterday because of the DREAMS."

The interview was concluded by providing participants the opportunity to provide open ended feedback. We asked the participants to "Describe in your own words what changes you would make to improve this experience."